\documentclass[conference]{IEEEtran}
\IEEEoverridecommandlockouts
\usepackage{cite}
\usepackage{amsmath,amssymb,amsfonts}
\usepackage{algorithmic}
\usepackage{graphicx}
\usepackage{textcomp}
\usepackage{xcolor}
\usepackage{booktabs}
\usepackage{todonotes}
\usepackage{comment}

\usepackage[utf8x]{inputenc}
\usepackage{float}

\def\BibTeX{{\rm B\kern-.05em{\sc i\kern-.025em b}\kern-.08em
    T\kern-.1667em\lower.7ex\hbox{E}\kern-.125emX}}
\begin{document}

\title{The Influence of Audio on Video Memorability with an Audio Gestalt Regulated Video Memorability System\\

\thanks{This work was partly supported by Science Foundation Ireland (SFI) under Grant Number SFI/12/RC/2289\_P2, co-funded by the European Regional Development Fund.}
}

\author{\IEEEauthorblockN{Lorin Sweeney}
\IEEEauthorblockA{\textit{Insight Centre for Data Analytics} \\
\textit{Dublin City University}\\
Dublin, Ireland \\
lorin.sweeney8@mail.dcu.ie}
\and
\IEEEauthorblockN{Graham Healy}
\IEEEauthorblockA{\textit{Insight Centre for Data Analytics} \\
\textit{Dublin City University}\\
Dublin, Ireland \\
graham.healy@dcu.ie}
\and
\IEEEauthorblockN{Alan F. Smeaton}
\IEEEauthorblockA{\textit{Insight Centre for Data Analytics} \\
\textit{Dublin City University}\\
Dublin, Ireland \\
alan.smeaton@dcu.ie}
}

\maketitle

\begin{abstract}
Memories are the tethering threads that tie us to the world, and memorability is the measure of their tensile strength. The threads of memory are spun from fibres of many modalities, obscuring the contribution of a single fibre to a thread's overall tensile strength. Unfurling these fibres is the key to understanding the nature of their interaction, and how we can ultimately create more meaningful media content. In this paper, we examine the influence of audio on video recognition memorability, finding evidence to suggest that it can facilitate overall video recognition memorability rich in high-level (gestalt) audio features. We introduce a novel multimodal deep learning-based late-fusion system that uses audio gestalt to estimate the influence of a given video's audio on its overall short-term recognition memorability, and selectively leverages audio features to make a prediction accordingly. We benchmark our audio gestalt based system on the Memento10k short-term video memorability dataset, achieving top-2 state-of-the-art results.
\end{abstract}

\begin{IEEEkeywords}
Memorability, multimodal, audio gestalt, deep learning
\end{IEEEkeywords}

\section{Introduction and Related Work}

Memories are the keepers of our continuity of self---without them, the very fabric of our being would fray. Yet, we scarcely have any influence on what we will ultimately remember or forget. The brain presides over the mechanisms of our memory from an opaque glass office, exercising sole editorial influence over its edifice. In fact, our odds of guessing what we will remember aren't much better than chance \cite{photograph2013memorable}. This lack of meta-cognitive insight, which prevents us from diving into our unconscious undercurrent, is what motivates and brings meaning to the exploration of the determinants of memory, and more specifically memorability---generally known as the likelihood that something will be remembered or forgotten. Naturally, the quantification of memorability is dependent on how we measure remembrance---which is in turn dependent on the modality and measurement paradigm.

Broadly speaking, there are two ways to measure remembrance: as \textit{recognition}, where amidst content presentation, participants indicate which items they feel they have previously perceived; or as \textit{recall}, where participants recount as much information as they can concerning previously presented content. These two measures respectively align with the two memory processes posited by the psychological dual process model of memory called \textit{process dissociation} \cite{jacoby1991process}. The first memory process is rapid, unconscious, and driven by a feeling of familiarity while the other is slower, conscious, and driven by a detail retrieving intention.

The three most common modalities with which memorability is explored are visual, textual, and auditory.
\subsection{Visual Memorability}

The predominant visual memorability measurement paradigm is \textit{recognition}---where memorability is commonly defined as the percentage of correctly recognised targets \cite{khosla2015understanding}. 
Using this paradigm, a high degree of human consistency concerning which images are remembered or forgotten is observed, suggesting that ``recognition memorability" is an intrinsic property of an image. Recognition memorability is also robust for other types of items, such as abstract visualisations \cite{borkin2013makes}, and specific objects within scenes \cite{dubey2015makes}. This intrinsic property is  not limited to static images, with faces shown to be consistently memorable across expression and viewpoint distortions \cite{bainbridge2017memorability}, and videos shown to be highly consistent in memory performance for both soundless 10-second movie clips \cite{cohendet2018annotating}, and 3-6 second viral videos \cite{mediaeval2020memory}. 

These results suggest that recognition memorability may be an intrinsic attribute of a wide range of stimulus types, even those with very different visual and semantic structures. Accordingly, relating it to other well-characterised image properties, such as saliency; colour features; aesthetics; etc., has been an active area of research. While several characteristics that correlate with recognition memorability have been proposed, a fully defining combination of features has yet to be identified. Simple image features, such as hue; saturation; or spatial frequency, have repeatedly been found not to correlate with recognition memorability \cite{bainbridge2017memorability, dubey2015makes, photograph2013memorable}. The number of objects depicted in an image does not appear to directly relate to its overall recognition memorability \cite{image2011memorable}, and likewise with properties such as aesthetics and interestingness \cite{photograph2013memorable}. However, combinations of semantically based attributes, such as object/scene category, emotion or actions, are predictive of recognition memorability \cite{cohendet2019videomem}. Additionally, scrambled images retain consistencies in recognition memorability, but only for short time periods (seconds). These findings suggest that recognition memorability is more closely linked to high-level perceptual properties of an image rather than low-level visual properties.

\subsection{Textual Memorability}

Many basics visual recall memorability findings---recall as a function of serial presentation position---are also observed in the textual equivalent \cite{deese1957serial}. However, repeated recall has been found to incrementally increase subsequent recall performance for images, but not for words \cite{erdelyi1974hypermnesia}. Words that arouse stronger emotion and are easier to visualise, exhibit enhanced recall \cite{bock1986influence}, while concrete words that refer to things that can be experienced by the senses, have a relative advantage in recall over abstract words. Words with smaller sets of associated words have an advantage over those with larger sets \cite{nelson1992word}. Minimally counter-intuitive concepts have been found to lead to better recall, suggesting that recall memorability is not an inherent property of a concept, but a property of the concept in the context it is presented \cite{upala2007contextualizing}.

Similar to images, the recognition memorability of simple words is highly consistent across individuals, suggesting that it is  an intrinsic property of words \cite{mahowald2018memorable}. Less familiar, lower-frequency words \cite{garlock2001age}; imageable and concrete words \cite{klaver2005word},  emotionally salient words \cite{kensinger2003memory} and the semantic context \cite{jacoby1981relationship} in which they are presented, all enhance recognition memorability. Additionally, meanings of words are retained in favour of their lexical properties \cite{begg1974retention}.

\subsection{Auditory Memorability}

Research into audio recall memorability shows that naming or verbalising sounds (phonological-articulation) can improve recall \cite{bartlett1977remembering}, and accordingly, non-verbal sounds have lower recall than verbal sounds \cite{paivio1975free}. Emotionality is known to play an important role in memory formation, and the emotional impact of a sound is correlated with the clarity of its perceived source \cite{HCU400}.
Human activity is considered to be a positively valenced sound \cite{dubois2006cognitive}, and positive valence improves sound recall \cite{jancke2008music}. It is generally accepted that auditory recall memorability is inferior to visual recall memorability, and decays more quickly \cite{bigelow2014}. However, it is important not to  overlook the role of the audio modality when exploring multi-modal media memorability, as multi-sensory experiences exhibit increased recall accuracy compared to uni-sensory ones \cite{thelen2015}, and sounds have the potential to provide valuable contextual priming information \cite{schirmer2011}.

While  little research has been conducted on auditory recognition memorability, interest has started to grow. Recent research suggests that similar to images and words, recognition memorability is an intrinsic property of sounds \cite{soundMem}.

\subsection{The MediaEval Memorability Task}

The MediaEval2020 memorability task \cite{mediaeval2020memory} is an annual event which benchmarks the effectiveness of predicting video memorability automatically.  In 2020 this operated on video data which included audio for the  first time, and several participants included audio features in their approaches to computing video memorability. Our approach and submission to the benchmark included audio gestalt features and produced promising preliminary results on the development test set of videos, shown in Table~\ref{tab:results1}.  Due to the abnormally low, participant-wide results on the official validation set shown in Table~\ref{tab:results2}, and an omission in our official submissions, very little insight into the efficacy of our approach was gained. However, one of our audio-based submissions \cite{sweeney2020leveraging} did achieve the best-in-class results from among all participants for long-term memorability predictions.

\begin{table}[H]
\caption{MediaEval2020 Media Memorability task results on 200 dev-set videos kept for validation for each of our runs.}
\label{tab:results1}
\begin{tabular}{cccc}
    \toprule
     &\multicolumn{1}{c}{Short-term} && \multicolumn{1}{c}{Long-term} \\\cline{2-4}
    \textbf{Run} & \textbf{Spearman}     && \textbf{Spearman}  \\
    Aug Captions + Spectrogram & 0.345 && 0.365  \\
    Captions + Frames & 0.338  && 0.437  \\
    Everything & 0.319 && 0.425  \\
    Audio Gestalt Spectrogram & \textbf{0.364}   && \textbf{0.470}  \\
    memento10k & 0.314  && -  \\
    \bottomrule
\end{tabular}
\label{tab:tab3}
\end{table}

\begin{table}[H]
\caption{Official MediaEval2020 Media Memorability task results on test-set for each of our submitted runs.}
\label{tab:results2}
\begin{tabular}{cccc}
    \toprule
     &\multicolumn{1}{c}{Short-term} && \multicolumn{1}{c}{Long-term} \\\cline{2-4}
    \textbf{Run} & \textbf{Spearman}     && \textbf{Spearman}  \\
    Aug Captions + Spectrogram & 0.054 && \textbf{0.113}  \\
    Captions + Frames & 0.05  && 0.059  \\
    Everything & - && 0.119  \\
    Audio Gestalt Spectrogram & 0.076   && 0.041 \\
    memento10k & \textbf{0.137}  && - \\
    \bottomrule
\end{tabular}
\label{tab:tab4}
\end{table}

In this paper, we evaluate the utility of including the audio modality in short-term video ``recognition memorability" prediction, and assess our gestalt based video memorability prediction system by benchmarking it on the Memento10k dataset \cite{mem10k}, comparing it to state-of-the-art solutions. Our contributions are two-fold: A) we assess the influence of the audio modality on video memorability B) we propose a multimodal deep learning-based late fusion system that uses audio gestalt to estimate the influence of the audio modality on overall video memorability, and selectively leverage audio features accordingly. Due to the nature of the Memento10k dataset, the recognition memorability in question is short-term.

\section{Methodology}

\subsection{Audio Gestalt Regulated Video Memorability}

\begin{figure}[H]
\includegraphics[width=1\columnwidth]{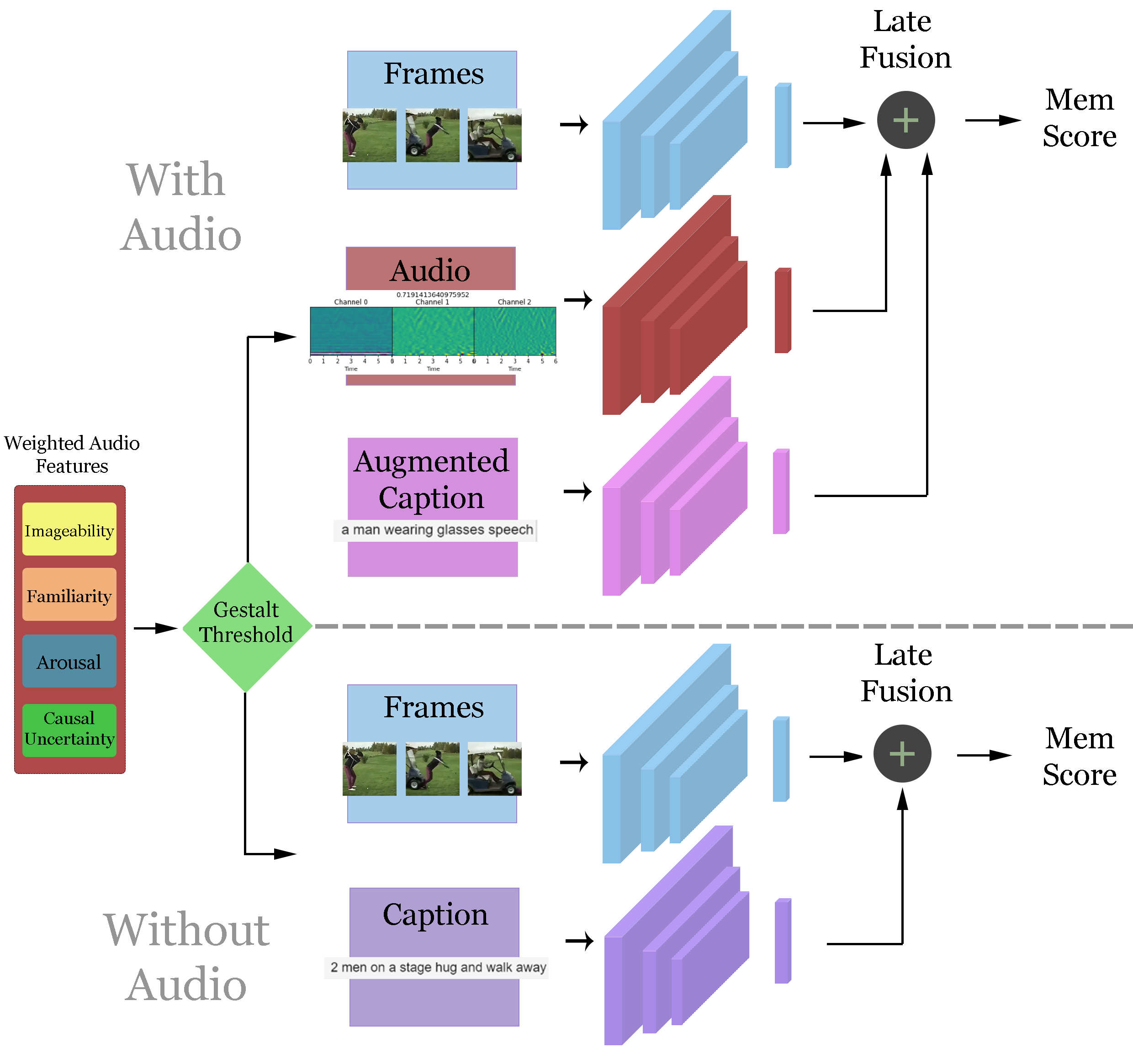}
\caption{Our multimodal deep-learning based late fusion framework,  using a conditional audio gestalt based threshold.}
\label{fig:framework}
\end{figure}

Our system is a multimodal deep-learning based late fusion framework that uses an audio gestalt conditional mechanism to predict short-term video recognition memorability Figure~\ref{fig:framework}. Depending on an audio gestalt threshold (0.8), one of two pathways---\textit{without audio}, using textual and visual features; and \textit{with audio}, using textual, visual, and auditory features---is used to predict a video's recognition memorability score. The \textit{without audio} stream's predictions are the weighted sum of our \textit{Frame} model (0.38), and \textit{Caption} model (0.62), while the \textit{with audio} stream's predictions are the weighted sum of our \textit{Frame} model (0.4), \textit{Augmented Caption} model (0.47), and \textit{Spectrogram} model (0.13). Both the weightings of the models's predictions and the gestalt threshold are determined using Randomised Search Cross-Validation (RSCV) from 0 to 1, in increments of 0.01.

\subsection{Audio Gestalt}
The Gestalt principles were first introduced by \cite{wertheimer1938laws} in 1928, and continue to be relevant in modern psychology. Traditionally thought of as rules that characterise the organisation of visual scenes---helping us understand them better---the Gestalt principles of \textit{similarity}; \textit{connectedness}; \textit{common region}; \textit{spatial proximity} \cite{gestalt2013memory}, and \textit{goodness} \cite{goetschalckx2019get} have been shown to benefit visual recognition memorability.

The very first usage of the term Gestalt was in 1890 in \cite{gestalt1890original}, which observed that humans can recognise two identical melodies even when no two corresponding notes have the same frequency. It was suggested that this property indicated the presence of a ``Gestalt quality"---a conceptual characteristic that assists our ``big picture" understanding of complex sensory data composed of many different parts. Unfortunately, since then, few insights intersecting audio gestalt and other well established audio properties have been revealed. The concept of gestalt in the context of audio was recently reintroduced by \cite{soundMem}, using the term gestalt to encapsulate high-level conceptual audio features. They found the following gestalt features: imageability; human causal uncertainty (Hcu); arousal; and familiarity, to be strongly correlated with audio memorability. In in this paper, we aim to practically apply these findings with the goal of elucidating the role of audio in overall video recognition memorability.

We create our own audio gestalt predictor using a weighted sum of our proxy measures for these four features. RSCV between 0 and 1 in increments of 0.05 is used to determine each of the weights. Due to the strong negative correlation between sound imageability and musicality \cite{bowles2016pitch}, we predicate imageability on whether the audio is classified as music or not. We use the PANNs \cite{panns} network to generate audio-tags, labelling the audio as music (giving it a score of 1.0) if a musical tag is present in the top 75\% confidence. Hcu and arousal scores are independently predicted with ImageNet-pretrained xResNet34 models fine-tuned on spectrograms from the HCU400 dataset \cite{HCU400}. Due limited available options, for familiarity, we use the top audio-tag confidence score of the PANNs \cite{panns} network as a proxy (Spearman = 0.305, pval = 4.749e-10 between the two scores in the HCU400 dataset). These four scores are then normalised (scaled into a 0-1 range), and a weighted score (with weights of 0.2, 0.2, 0.2, and 0.4 respectively) is calculated to produce an audio gestalt score.

\begin{figure}[ht]
\includegraphics[width=1\columnwidth]{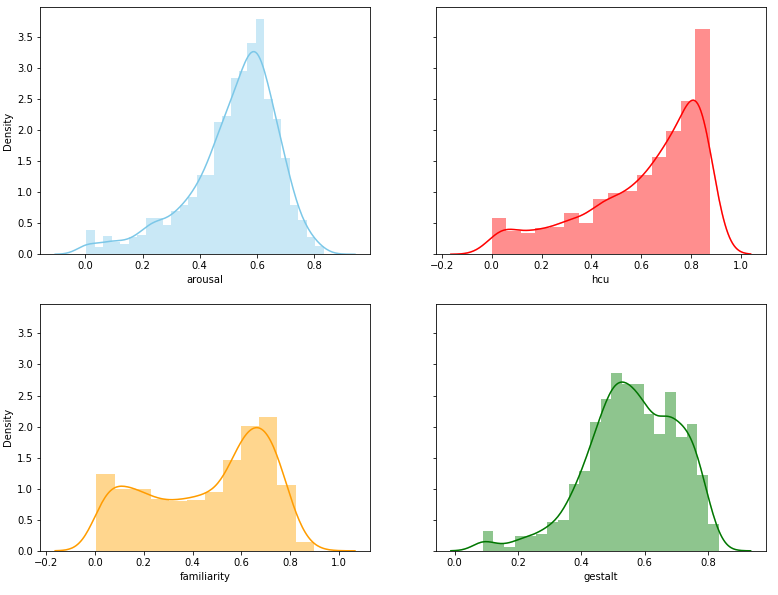}
\caption{Distribution of audio gestalt and gestalt related audio features from 1,468 validation videos.}
\label{fig:distribution}
\end{figure}

\begin{figure}[ht]
\includegraphics[width=1\columnwidth]{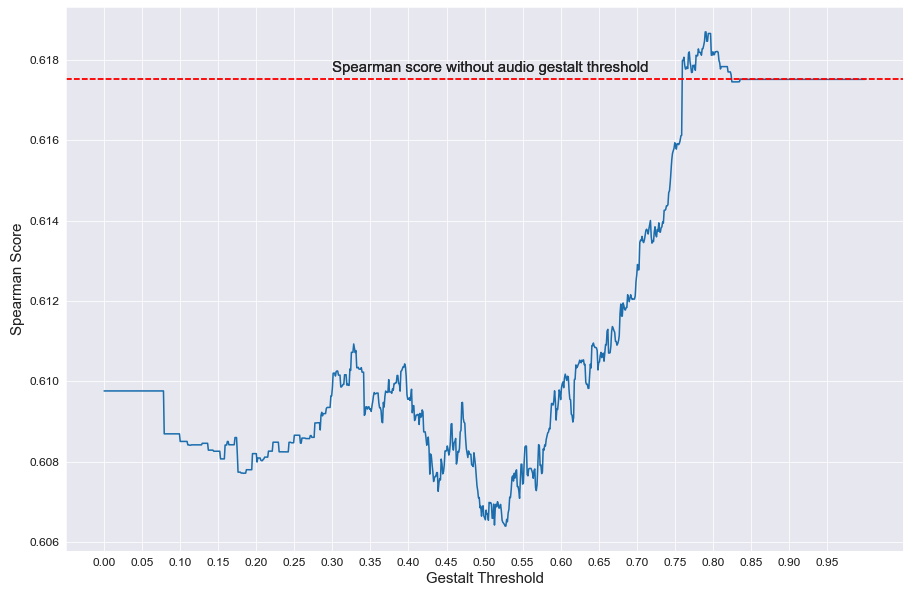}
\caption{Effect of gestalt thresholds on Spearman scores of 1,468 Memento10k validation videos.}
\label{fig:threshold}
\end{figure}

\subsection{Auditory Features}

For auditory features, we train a network to predict a video's recognition memorability from audio spectrograms---our \textit{Spectrogram model}. We extracted Mel-frequency cepstral coefficients (n\_fft:2048, hop\_length:256, n\_mels:128) from the the 6,890 Memento10k \cite{mem10k} training videos with audio, and stacked them with their delta coefficients in order to create three channel spectrogram images. These spectrogram images are then used to train an ImageNet-pretrained xResNet34 model for 15 epochs; with a max learning rate of 1e-2; and weight decay of 1e-3 to predict audio recognition memorability. Additionally, we fit a Bayesian Ridge Regressor with VGGish \cite{vggish} audio features---our \textit{Bayesian Ridge} model. We extract 128-dimensional embeddings for each second of video audio, resulting in a 384-dimensional feature set per video.

\subsection{Visual Features}
We evaluate the extent to which static visual features contribute to video recognition memorability by training a network to predict a video’s recognition memorability from a single frame---our \textit{Frame} model. We train an ImageNet-pretrained xResNet50 to predict image recognition memorability by first training on the LaMem dataset \cite{khosla2015understanding} for 50 epochs; with a maximum learning rate of 3e-2; and weight decay of 1e-2, and then fine-tuning on the 6,890 Memento10k \cite{mem10k} training videos which have audio with the same hyperparameters. At test time, a video’s recognition memorability score is calculated by averaging predictions of the first, middle, and last frame. 

\subsection{Textual Features}

For textual features, we train a network to predict a video's recognition memorability from a paragraph of text composed of five independently generated human captions---our \textit{Caption model}. Given that overfitting is a primary concern, we use the AWD-LSTM (ASGD Weight-Dropped LSTM) architecture \cite{AWD_LSTM}, as it is highly regularised, and is comparable to other state-of-the-art language models. In order to fully take advantage of the high level representations that a language model offers, we transfer train our model using UMLFiT \cite{UMLFit}, a method that uses discriminative fine-tuning, slanted triangular learning rates, and gradual unfreezing to avoid catastrophic forgetting. 

A Wiki-103-pretrained language model is fine-tuned on the first 300,000 captions from Google's Conceptual Captions dataset \cite{GCC} for a total of 10 epochs with a dropout multiplier of 0.5 and max learning rate of 2e-3, resulting in a final language model accuracy of 37\%. The encoder from that model is re-used in another model of the same architecture, but trained on captions from the 6,890 Memento10k \cite{mem10k} training videos with audio, for a total of 15 epochs with a dropout multiplier of 0.8 and a max learning rate of 1e-3, to predict recognition memorability scores, rather than the next word in a sentence. An additional network is trained the same way, but fine-tuned on captions that are augmented with audio tags extracted using the PANNs \cite{panns} network---our \textit{Augmented Caption model}.

In all cases, models were independently trained on the 6,890 Memento10k training set videos with audio, and independently validated on the 1,484 Memento10k validation videos with audio. All parameter tuning (e.g. RSCV) was performed using the Memento10K training set.

\section{Results}

As in the MediaEval memorability task, prediction performance is measured by calculating the Spearman’s rank correlation of the predicted memorability rankings with their ground truth rankings. Table~\ref{tab:results3} shows the Spearman rank correlation scores of the individual components of our audio gestalt system, many of their combinations, and the final implementation of our audio gestalt system on the 1,484 Memento10k validation videos with audio. The best performing individual component is our \textit{Caption model}, achieving a Spearman score of 0.5710. Each of the component combinations are the result of a randomised search weighted summation of their predictions, with the best combination being \textit{Captions + Frames} (0.6175). Our audio gestalt based system was the best performing approach, achieving a Spearman score of 0.6181.

\begin{table}[ht]
\caption{Results on 1,484 memento10k validation videos with audio.}
\label{tab:results3}
\begin{tabular}{llcc}
     &\multicolumn{1}{c}{Memorability} \\
     \toprule
    \textbf{Approach} & \textbf{Spearman} \\
    \midrule
    Spectrogram & 0.2030 \\
    Bayesian Ridge & 0.2913 \\
    Frames & 0.4808 \\
    Frames + Spectrogram & 0.4876 \\
    Frames + Bayesian Ridge & 0.4992 \\
    Captions & 0.5710 \\
    Captions + Spectrogram & 0.5715 \\
    Captions + Bayesian Ridge & 0.5741 \\
    Augmented Captions & 0.5555 \\
    Augmented Captions + Spectrogram & 0.5562  \\
    Augmented Captions + Bayesian Ridge & 0.5576  \\
    Augmented Captions + Frames & 0.6068 \\
    Captions + Frames & 0.6175 \\
    Everything Ridge & 0.6066 \\
    Everything Spectrogram & 0.6061 \\
    Audio Gestalt Ridge Normal Captions & 0.6175 \\
    Audio Gestalt Spectrogram Normal Captions & 0.6176 \\
    Audio Gestalt Ridge & 0.6181 \\
    Audio Gestalt Spectrogram & \textbf{0.6181} \\ \bottomrule
\end{tabular}
\end{table}

To evaluate the effectiveness of our approach, we compare against the Memento10k benchmark scores \cite{mem10k}. From Table~\ref{tab:results4} we can see that our audio gestalt based approach outperforms all other approaches except SemanticMemNet \cite{mem10k}---the model introduced alongside the Memento10k dataset.

\begin{table}[ht]
\caption{Comparison of state-of-the-art on Memento10k. *Trained and validated on fewer videos due to audio constraint, 7,000 vs. 6,890 and 1,500 vs. 1484 respectively.}
\label{tab:results4}
\begin{tabular}{llcc}
     &\multicolumn{1}{c}{Memorability} \\
     \toprule
    \textbf{Approach} & \textbf{Spearman} \\
    \midrule
    Human Consistency & 0.730 \\
    \hline
    MemNet Baseline \cite{khosla2015understanding} & 0.485 \\
    Cohendet et al. (Semantic) \cite{cohendet2019videomem} & 0.552 \\
    Cohendet et al. (ResNet3D) \cite{cohendet2019videomem} & 0.574 \\
    Feature Extraction + Regression (as in \cite{shekhar2017show}) & 0.615 \\
    SemanticMemNet \cite{mem10k} & 0.663 \\
    \hline
    Audio Gestalt & \textbf{0.618}* \\ \bottomrule
\end{tabular}
\end{table}

With respect to our results in Table ~\ref{tab:results3}, the general trend for predicting video recognition memorability seems to be that the more modalities used, the better the predictions. Even the addition of a poorly-performing individual audio model (0.2913) with a better-performing individual visual model (0.4808), produces an increase in performance (0.4992). There are however, some very important exceptions to this trend. Indiscriminately tri-modal approaches, \textit{Everything Ridge} (0.6066) and \textit{Everything Spectrogram} (0.6061), achieve lower Spearman scores than the bi-modal combination of visual and textual predictions (0.6175), and their selectively tri-modal counterparts (0.6181). 

At first glace, it appears that augmenting captions with audio-tags is worse than vanilla captions, Augmented Captions (0.5555) vs Captions (0.5710); Augmented Captions + Spectrogram (0.5562) vs Captions + Spectrogram (0.5715); Augmented Captions + Bayesian Ridge (0.5576) vs Captions + Bayesian Ridge (0.5741); Augmented Captions + Frames (0.6068) vs Captions + Frames (0.6175), however, when selectively used in our audio gestalt system (0.6181), they outperform vanilla captions (0.6175).

\section{Discussion}

Our audio gestalt based system ultimately outperforms all of our other tested approaches. Even though the advantage incurred is only marginal, selectively including audio features (0.6181) is ultimately better than both always including them (0.6066), and not including them (0.6175).  We believe that this can in part be explained by the fact that sounds have the have the potential to provide valuable contextual priming information \cite{schirmer2011}, but that some sounds simply add noise, having a deleterious effect on overall understanding of a context. Thinking of audio gestalt as an ontological property that encapsulates high-level auditory features that positively contribute towards our understanding of a context, helps explain the benefit of using it as a measure to discriminate between useful and distracting audio in multimodal content. The effect of different gestalt thresholds is shown in Figure 3.

It is interesting to note that there is no difference in Spearman score between Audio Gestalt Spectrogram (0.6181) and Audio Gestalt Ridge (0.6181), even though the Bayesian Ridge achieves a noticeably higher Spearman score (0.2913) than the Spectrogram model (0.2030). This indicates that the inclusion of auditory features is not strictly additive, and further suggests that they may act as a contextual signal of some sort.

In \cite{soundMem}, they found that the strongest predictors of sound recognition memorability were imageability, and causal uncertainty (Hcu). Naturally, we would expect our audio gestalt weightings to reflect this to some degree, but we found that the highest weighted audio gestalt feature is familiarity (top audio-tag confidence score). The gestalt weightings for imageability; Hcu; familiarity; and arousal, are 0.2; 0.2; 0.4; 0.2 respectively. As shown in Figure 2, familiarity is the only audio gestalt feature with a bi-modal distribution. Both arousal and Hcu are heavily left skewed, leading us to believe that the models used to predict their scores have been overfit, and can be improved.

\section{Conclusions}
In this paper we have assessed the influence of the audio modality on video recognition memorability, finding evidence to suggest that it primarily plays a contextualising role, with the potential to act as a signal or trigger that aids recognition depending on the extent of its high-level features. We introduced a novel multimodal deep learning-based late-fusion system that uses audio gestalt to estimate the influence of a given video's audio on its overall short-term recognition memorability that selectively leverages audio features to make a prediction accordingly. Our findings add further credibility to the hypothesis that recognition memorability is more closely linked to high-level perceptual properties of content than low-level properties, and that this relationship extends beyond the visual domain. Similar to the way in which textual memorability, both recall and recognition, has been suggested as not an inherent property of a concept, but a property of the concept in the context it is presented, the influence of auditory memorability in a multimodal medium such as video, is likely to be highly context dependent.

While this work has made progress towards understanding the influence of the audio modality on short-term video recognition memorability, the full extent of its role is far from being understood. It is possible that the correlation between the content/context of a video's auditory modality and its visual modality could play an important role in determining the audio's impact on the video's overall recognition memorability, however, without testing this experimentally, we simply cannot answer this question. We believe that improvements can be made by refining our measure of audio gestalt.

Independent memorability scores for each of the modalities---audio, visual, and textual---would assist us in elucidating the role they each play when coinciding with one another in a multimodal medium such as video, and should be a focus of future memorability research. Similarly, recognition memorability and recall memorability would each benefit from a directed disentanglement effort as they are often conflated. The way in which they interact is relatively unexplored, and further study here is likely to yield valuable insights. 

\bibliographystyle{./bibliography/IEEEtran}
\bibliography{./bibliography/lorins_big_bib}

\end{document}